\documentclass[dvipdfm]{article}

\usepackage{amsmath, amsthm}
\usepackage{amsmath, amsfonts}

\newtheorem{thm}{Theorem}[section]

\newtheorem{prpt}[thm]{Property}
\numberwithin{equation}{section}

\newcommand{\pa}{\partial}
\newcommand{\pf}{\noindent \textbf{Proof}}
\newcommand{\ed}{\hspace*{10pt} \hfill $\square$}

\begin{document}

  %封面内容
  \title{ Supersymmtric Quantum Mechanics and Lefschetz fixed-point formula}
  \author{\hspace*{-20pt}Si\ \hspace*{5pt} Li
  \\{\it \normalsize {$^1$ Department of Mathematics},}\ \ {\it \normalsize {University of Science and Technology of
  China }  }\\\it{\normalsize{Hefei, Anhui, 230026, P.R.China} }
  \\ $^2$ {\it \normalsize{ USTC Shanghai Institute for Advanced Studies}}
  \\ \it{\normalsize{Shanghai, 201315,\ P.R.China}}
  \\
  \\ Email: lisi@mail.ustc.edu.cn}
 \date{}
  % 封面
  \maketitle

%%%%%%%%%%%%%%%%%%%%%%%%%%%%%%
%% 前言部分
%%%%%%%%%%%%%%%%%%%%%%%%%%%%%%

\begin{abstract}
We review the explicit derivation of the Gauss-Bonet and Hirzebruch
formulae by physical model and give a physical proof of the
Lefschetz fixed-point formula by twisting boundary conditions for
the path integral.

\end{abstract}

  % 目录
%  \tableofcontents
  % 表格目录
%  \listoftables
  % 插图目录
%  \listoffigures

%%%%%%%%%%%%%%%%%%%%%%%%%%%%%%
%% 正文部分
%%%%%%%%%%%%%%%%%%%%%%%%%%%%%%
%\mainmatter
\section{Introduction}
Physical models with supersymmetry are known to have close relation
with mathematical structures. The Atiyah-Singer formula\cite{Atiyah}
for the index of Dirac operators as well as the index formulae for
Euler number, signature, Todd genus, Hirzebruch $\chi_y$-genus can
be derived from a simple supersymmetric quantum mechanical model
\cite{AG,FW,Meng}.\\
\\
In this paper we show how to explicitly derive the Gauss-Bonet ,
Hirzebruch and Lefschetz fixed-point formulae from the
supersymmetric sigma model and point out that the choices of
boundary conditions in the path are crucial to the path integral
calculation and can be realized as classical operators on the
geometric objects. A special and unfamiliar boundary condition is
chosen in order to derive the Lefschetz fixed-point formula. \\
\\
The organization of the paper is as follows: In Sect.2, we give a
quick review of supersymmetric sigma model on Riemannian manifold
that we will use throughout this paper. In Sect.3 we use this model
to derive Gauss-Bonet, Hirzebruch and Leftschetz fixed-point
formulae.

\section{Supersymmetric Quantum Mechanics on Riemannian manifold}
We give a short review of Witten's definition of Supersymmetric
Quantum Mechanics (SQM) in the Appendix. In this section, we study a
simple but important example of SQM defined on a Riemannian
manifold. For more details
consult \cite{mirror symmetry}.\\
\\
Let $\left(M,g\right)$ be an oriented and compact Riemannian
Manifold of dimention n with metric g. $I=[0,T]$ is the time
interval. The bosonic field defines a map
\begin{eqnarray}
    \phi:I\longrightarrow M,
\end{eqnarray}
which is represented locally as $x^I\circ \phi =\phi^I$.
 Here $\{x^I\},I=1,\ldots,n$, is the local coordinate on M. The fermionic
variables define sections
\begin{eqnarray}
    \psi,\bar{\psi}\in \Gamma(I,\phi^*TM\otimes C),
\end{eqnarray}

We can construct a Lagrangian for the bosonic and fermionic fields
\begin{eqnarray}\label{riem}
L={1 \over 2}g_{IJ}\dot{\phi}^I \dot{\phi}^J +{i \over
2}g_{IJ}(\phi)\left(\bar{\psi}^I D_t \psi^J-D_t \bar{\psi}^I \psi^J
\right)-{1 \over 4} R_{IJKL}\psi^I \psi^J \bar{\psi}^K \bar{\psi}^L
\end{eqnarray}
where $D_t \psi^I={\pa\over \pa t}\psi^I
+\Gamma^I_{JK}\dot{\phi}^J\psi^K$ and the summation convention is
used. This lagrangian preserves the SUSY transformation
\begin{eqnarray}
    &&\delta \phi^I=\epsilon \bar{\psi}^I-\bar{\epsilon} \psi^I\\
    &&\delta \psi^I=\epsilon \left(i\dot{\phi}^I -\Gamma^I_{JK} \bar{\psi}^J \psi^K
    \right)\\
    &&\delta \bar{\psi}^I=\bar{\epsilon} \left( -i\dot{\phi}^I-\Gamma^I_{JK} \bar{\psi}^J\psi^K \right)
\end{eqnarray}
The supersymmetry charges are given by
\begin{eqnarray}
    &&Q=ig_{IJ}\bar{\psi}^I \dot{\phi}^J=i\bar{\psi}^I P_I\\
    &&\bar{Q}=-ig_{IJ}\psi^I \dot{\phi}^J=-i\psi^I P_I
\end{eqnarray}
where $P_I=g_{IJ}\dot{\phi}^J$ is the conjugate momentum of
$\phi^I$.
\\

After the canonical quantization ,we will have
\begin{eqnarray}
    &&\left[ \phi^I,P_J \right]=i\delta^I_J\\
    &&\left\{ \psi^I,\bar{\psi}^J \right\}=g^{IJ}
\end{eqnarray}
the Fermion number operator is
\begin{eqnarray}
    F=g_{IJ}\bar{\psi}^I \psi^J
\end{eqnarray}
\\
The natural Hilbert space equipped with this quantum mechanical
system is
\begin{eqnarray}
    {\it F}=\Omega^*(M)\otimes C
\end{eqnarray}
with the Hermitian inner product
\begin{eqnarray}
    (\omega_1,\omega_2)=\int_M \bar{\omega}_1 \wedge \ast\omega_2
\end{eqnarray}

The obserbables are represented on this Hilbert space as familiar
operators
\begin{eqnarray*}
    &&\phi^I=x^I \times\\
    &&P_I =-i\nabla_I\\
    &&\bar{\psi}^I=dx^I\wedge\\
    &&\psi^I=g^{IJ}i_{{\pa \over \pa x^J}}
\end{eqnarray*}
Denote the vacuum $|0>$ satisfying $\psi^I |0>=0,\forall I$, then
\begin{eqnarray*}
    |0>&\leftrightarrow& 1\\
    \bar{\psi}^I |0>&\leftrightarrow& dx^I\\
    &\vdots&\\
    \bar{\psi}^1\cdots \bar{\psi}^n|0>&\leftrightarrow& dx^1\wedge
    \cdots dx^n\\
    Q=i\bar{\psi}^I P_I &\leftrightarrow& d=dx^I\wedge \nabla_I\\
    Q^\dag=-i\psi^I P_I &\leftrightarrow& d^\dag=-g^{IJ} i_{\pa \over \pa
    x^J} \wedge {\pa \over \pa x^I}\\
    H={1 \over 2}\left\{ Q,\bar{Q} \right\}&\leftrightarrow& {1 \over
    2} \triangle ={1 \over 2} \left\{ dd^\dag+d^\dag d \right\}
\end{eqnarray*}

The supersymmetric ground states are obviously given by
\begin{eqnarray}
H_{(0)}=H(M,g)=\bigoplus\limits^n_{p=0}H^p(M,g)
\end{eqnarray}
where $H(M,g)$ is the space of harmonic forms of the Riemannian
manifold $(M, g)$ and $H^p(M, g)$ is the space of harmonic p-forms.\\
\\
On the other hand we know that the space of supersymmtric ground
states is isomorphic to the Q-cohomology  which here is represented
by the De Rham cohomology. Take care of the grading by fermion
number we will have
\begin{eqnarray}
    H^p(Q)=H^p_{DR}(M)
\end{eqnarray}
We get the famous formula
\begin{eqnarray}
    H^p(M,g)\cong H^p_{DR}(M)
\end{eqnarray}
The supersymmetric index is the Euler characteristic of the
Q-complex, namely
\begin{eqnarray}\label{euler}
Tr(-1)^Fe^{-\beta H}=\sum\limits^n_{p=0}(-1)^p dim
H^p(Q)=\sum\limits^n_{p=0}(-1)^p dim H^p_{DR}(Q)=\chi(M)
\end{eqnarray}

\section{Mathematical application of SQM}
\subsection{Gauss-Bonnet foumula}
We consider the SQM model on a Riemannian manifold $(M,g)$ discussed
in section 2. The lagrangian is
\begin{eqnarray}
L={1 \over 2}g_{IJ}\dot{\phi}^I \dot{\phi}^J +i
g_{IJ}(\phi)\bar{\psi}^I D_t \psi^J-{1 \over 4} R_{IJKL}\psi^I
\psi^J \bar{\psi}^K \bar{\psi}^L
\end{eqnarray}
Note that we choose a convenient lagrangian here which differs from
eqs.(\ref{riem}) by a total derivative. After wick rotation
$t\rightarrow -it$, we get
\begin{eqnarray}\label{s euler}
    S_E&=&\int_0^{\beta} dt \ \left\{{1 \over 2} g_{IJ}\dot{\phi}^I
    \dot{\phi}^J+g_{IJ}\bar{\psi}^I D_t \psi^J +{1 \over 4} R_{IJKL}\psi^I \psi^J \bar{\psi}^K
    \bar{\psi}^L\right\}\label{p1}
\end{eqnarray}
Since the Witten index is independent of $\beta$, we take $\beta
\rightarrow 0$ and eqs.(\ref{s euler}) shows that the path integral
will localize on the constant maps. We can fourier expand the fields
near the constant maps
\begin{eqnarray}\label{expansion euler}
 &&\dot{\phi}^I=x^I_0+\sqrt{\beta} \sum\limits^{\infty}_{n\neq
 0}a^I_n e^{2\pi nti/\beta}\\
 &&\bar{\psi}^I=\beta^{1/4} \bar{\psi}^I_0+\sum\limits^{\infty}_{n\neq
 0}\bar{\psi}^I_0 e^{2\pi nti/\beta}\\
 &&\psi^I=\beta^{1/4}\psi^I_0+\sum\limits^{\infty}_{n\neq
 0}\psi^I_0 e^{2\pi nti/\beta}
\end{eqnarray}
The factors of $\sqrt{\beta}$ and $\beta^{1/4}$ in the mode
expansion is included to keep the first order approximation and
remove the $\beta$-dependence from the integration measure. We will
integrate out first all the nonzero modes of all fields and then all
zero modes. Since the path integral is invariant under change of
variables, we can use normal coordinates centered at the point where
we do the expansion. Take the above expansion back to eqs.(\ref{s
euler}), we get
\begin{eqnarray}
    S_E=\sum\limits_{n\neq 0} \left\{{(2\pi n)^2 \over 2}a^I_n (a^I_n)^*
    +(2\pi n i)\bar{\psi}^I_n \psi^I_n\right\}+{1 \over 4} R_{IJKL}(x^I_0)\psi^I_0 \psi^J_0
    \bar{\psi}^K_0
    \bar{\psi}^L_0+O(\beta)\ \
\end{eqnarray}
The integration of nonezero modes will give 1 which is in fact a
consequence of supersymmetry. The integration of zero modes gives
\begin{eqnarray*}
    \chi(M)=Tr(-1)^Fe^{-\beta H}={1 \over (2\pi)^{n/2}} \int d(Vol) \int
    \prod\limits_m d \psi^m_0 d\bar{\psi}^m_0 exp \left\{-{1 \over 4} R_{IJKL}(x^I_0)\psi^I_0 \psi^J_0
    \bar{\psi}^K_0
    \bar{\psi}^L_0 \right\}
\end{eqnarray*}
where n is the dimension of the manifold. If n is odd, the above
equation gives zero; if n is even ,the expression becomes
\begin{eqnarray*}
    \chi(M)&=&Tr(-1)^Fe^{-\beta H}\\
           &=&{(-1)^{n/2} \over 2^n \left( {n \over 2} \right)!\pi^{n/2}
           }\int d(Vol) \varepsilon^{I_1J_1\ldots
           I_mJ_m}\varepsilon^{K_1L_1\ldots
           K_mL_m}R_{I_1J_1K_1L_1}\cdots R_{I_mJ_mK_mL_m} ,\ n=2m,
\end{eqnarray*}
which is just the Gauss-Bonet formula \cite{chen}.

\subsection{Hirzebruch signature}
A path integral derivation of $\chi_y$-genus is given by
\cite{Meng}. We use the similar procedure as \cite{Meng} to give an
explicit calculation of Hirzebruch signature. Consider SQM on a
Riemannian manifold $(M,g)$
\begin{eqnarray}
    L={1 \over 2}g_{IJ}\dot{\phi}^I \dot{\phi}^J +i g_{IJ}(\phi)\bar{\psi}^I D_t \psi^J-{1 \over 4} R_{IJKL}\psi^I \psi^J \bar{\psi}^K \bar{\psi}^L
\end{eqnarray}
 Notice that the lagrangian
has a discrete symmetry $\bar{\psi}\leftrightarrow \psi $. Let the
operator $\Gamma $ implementing this symmetry. Recall that the hodge
star operator acts on differential forms as
\begin{eqnarray}
    * : \Omega^p M \longrightarrow \Omega ^{n-p} M
\end{eqnarray}
for  $\varphi= {1 \over p\ !} \sum\limits_{i_1,\ldots,i_p}
\varphi_{i_1, \ldots, i_p} dx^{i_1} \wedge \cdots \wedge dx^{i_p}$ ,
\begin{eqnarray}
    &&*\varphi={1 \over (n-p)!} \sum\limits_{i_1, \ldots, i_n} {1 \over
    p\ !} \varepsilon _{i_1, \ldots, i_n} \varphi ^{i_1, \ldots, i_p}
    dx^{i_{p+1}}\wedge \cdots \wedge dx^{i_n}\\
    \mbox{where}&&\nonumber\\
    &&\varphi ^{i_1, \ldots, i_p}=\sum\limits_{j1, \ldots, j_p}
    g^{i_1j_1}\cdots g^{i_pj_p} \varphi _{j_1, \ldots, j_p}
\end{eqnarray}
In particular , $*1=d(Vol)$ , where $d(Vol)$ is the volume form
\begin{eqnarray*}
    d(Vol)&=&\sqrt{det(g)} dx^1\wedge \cdots \wedge dx^n\\
      &=&{1 \over n!} \varepsilon_{i_1, \ldots, i_n} dx^{i_1}\wedge
      \cdots \wedge dx^{i_n}
\end{eqnarray*}
The first obserbation is that the operator $\Gamma$ we defined
exchanges $\psi$ and $\bar{\psi}$ and therefore sends the vacuum
state $|0>$ to the state which is annihilated by $\bar{\psi}$.
Represented on the differential forms, this state is nothing but the
volume form $d(Vol)$, so we have
\begin{eqnarray}
\Gamma |0> = \sqrt{det(g)} \bar{\psi}^1\cdots\bar{\psi}^n |0>
\end{eqnarray}
for the general state
\begin{eqnarray}
    \varphi ={1 \over p\ !} \varphi_{i_1,\ldots, i_p}
    \bar{\psi}^{i_1}\cdots \bar{\psi}^{i_p} |0>
\end{eqnarray}
the action by $\Gamma$ gives
\begin{eqnarray}
    \Gamma \varphi&=& {1\over p\ !} \varphi_{i_1,\ldots,i_p}
    \psi^{i_p,\ldots,i_1} d(Vol)\\
    &=&{1\over p\ !}\varphi_{i_1,\ldots,i_p}
    \psi^{i_p,\ldots,i_1} {1\over n!} \varepsilon_{j_1,\ldots,j_n}
    \bar{\psi}^{j_1}\cdots\bar{\psi}^{j_n}|0>\\
    &=&{1\over p\ !} \varphi_{i_1,\ldots,i_p} {1 \over
    (n-p)!}g^{i_1j_1}\cdots g^{i_pj_p}\varepsilon_{j_1,\ldots,j_n}
    \bar{\psi}^{j_{p+1}}\cdots\bar{\psi^{j_n}}|0>\\
    &=&*\varphi
\end{eqnarray}
So if represented on the differential forms , the operator $\Gamma$
is just the hodge star operator. As a consequence, the Hirzebruch
signature can be expressed by
\begin{eqnarray}
    \tau(M)=Tr\Gamma (-1)^Fe^{-\beta H}= n^{E=0}(\Gamma=+1)-n^{E=0}(\Gamma=-1)
\end{eqnarray}

\noindent We come to the path integral calculation of Hirzebruch
signature. To implement the discrete symmetry, we redefine the
fermionic field
\begin{eqnarray}
    \psi_+^I={1\over 2} \left(\bar{\psi}^I+\psi^I \right)\\
    \psi_-^I={1\over 2} \left(\bar{\psi}^I-\psi^I \right)
\end{eqnarray}
and the lagrangian takes the form
\begin{eqnarray}
    L={1 \over 2}g_{IJ}\dot{\phi}^I \dot{\phi}^J+i g_{IJ}\psi^I_+ D_t
    \psi^J_+ -i g_{IJ}\psi^I_- D_t \psi^J_--{1 \over 4} R_{IJKL}\psi_+^I \psi_+^J \psi_{\_}^K \psi_{\_}^L
\end{eqnarray}
the discrete symmetry is transformed to
\begin{eqnarray}
    &&\psi_+ \longleftrightarrow \psi_+\\
    &&\psi_- \longleftrightarrow -\psi_-
\end{eqnarray}
Now by a standard argument ,
\begin{eqnarray}
    Tr\Gamma(-1)^Fe^{-\beta H}=\int_{BC} D\phi D\psi_+ D\psi_- e^{-S_E}
\end{eqnarray}
the boundary condition is just
\begin{eqnarray}
    &&\phi(\beta)=\phi(0)\\
    &&\psi_+(\beta)=\psi_+(0)\\
    &&\psi_-(\beta)=-\psi_-(0)\label{BC}
\end{eqnarray}
After wick rotation $t\rightarrow -it$ ,
\begin{eqnarray}\label{s+-}
    S_E=\int_0^{\beta} dt{1 \over 2}g_{IJ}\dot{\phi}^I \dot{\phi}^J+ g_{IJ}\psi^I_+ D_t
    \psi^J_+ - g_{IJ}\psi^I_- D_t \psi^J_-+{1 \over 4} R_{IJKL}\psi_+^I \psi_+^J \psi_{\_}^K
    \psi_{\_}^L\ \ \ \
\end{eqnarray}
A subtle fact is that $\psi_{\_}$ has no constant mode by the
boundary conditon (\ref{BC}) so we should take care of the first
order to do the approximation. As before we do the fourier expansion
\begin{eqnarray}
    &&\phi^I=x_0^I+\sqrt{\beta} \sum\limits_{n\neq 0}^{\infty} a_n^I
    e^{2 \pi n i t/\beta}\\
    &&\psi_+^I=\sqrt{i\over 2\pi \beta} \psi_0^I+\sum\limits_{n\neq 0}^{\infty} \psi_n^I
    e^{2 \pi n i t/\beta}\\
    && \psi_{\_}^I={1\over 2}e^{-{i\over 2}\pi+\pi t i}\eta_0^I +\sum\limits_{n\neq 0}^{\infty} \eta_n^I
    e^{2 \pi n i t/\beta}
\end{eqnarray}
Choose the normal coordinate near $x_0^I$ , substitute this back to
eqs.(\ref{s+-}) , take care of the connection term in $D_t$, keep
the lowest order, we get
\begin{eqnarray*}
    S_E&=&{1\over 2} \sum\limits_{n\neq 0} (2\pi n)^2 a^I_n
    (a^I_n)^*+\sum\limits_{n\neq 0} (2\pi
    ni)\psi^I_{-n}\psi^I_n-\sum\limits_{n\neq 0} 2n\pi i \eta^I_{-n-1}
    \eta ^I_n\\
    &&-\sum\limits_{n\neq 0}{n\over 2} \Omega_{KL} (a^K_n)^* a^L_n+\sum\limits_{n\neq 0}\left({i\over 2\pi}
    \Omega_{KL}-\pi i \delta_{KL}\right)\eta^K_{-n-1}\eta^L_n\\
    &&+e^{{i\over 2}\pi}\left(-{i\over 4\pi}\Omega_{KL}+{\pi i\over 2} \delta_{KL}\right)
    \eta^K_{-1}\eta^L_0+O(\beta)
\end{eqnarray*}
where
\begin{eqnarray}
    \Omega_{KL}=R_{IJKL}\psi^I_0\psi^J_0
\end{eqnarray}
Integrate out $a_n,(a_n)^\dag$ gives
\begin{eqnarray}
    \prod\limits_{n>0} det^{-1}\left( (2\pi n)^2 I- n\Omega
    \right) det^{-1} \left( (2\pi n)^2 I+ n\Omega
    \right)\\
\end{eqnarray}
Integrate out $\psi_n, \eta_n$ gives
\begin{eqnarray*}
    &&\left\{\prod\limits_{n>0} det\left( (2\pi n)^2 I \right)\right\}
    det\left(e^{{i\over 2}\pi}\left({i\over  4\pi}
    \Omega-{\pi i\over 2} I\right)\right) \\
    &&\left\{\prod\limits_{n>0}
    det\left( 2\pi n i-\left( {i\over 2\pi}\Omega-\pi i I \right) \right)
    det\left( -2\pi n i-\left( {i\over 2\pi}\Omega-\pi i I \right) \right)\right\}
\end{eqnarray*}
Put it together , we have
\begin{eqnarray*}
    det\left(e^{{i\over 2}\pi}\left({i\over  4\pi}
    \Omega-{\pi i\over 2} I\right)\right)
    \prod\limits_{n>0} det^{-1}\left( I+\left({i\Omega/4\pi \over n\pi} \right)^2 \right)
    det\left(I+ \left( {i\over 4\pi}\Omega-{\pi i\over 2}I \right)^2\right)
\end{eqnarray*}
To evaluate the above expression , we assume that ${i\over
4\pi}\Omega$ has eigenvalues $\chi_I$ and use the formula
\begin{eqnarray}
    &&\sinh x=x \prod_{n>0}\left(1+\left( {x\over n\pi} \right)^2
    \right)\\
    &&\cosh x=e^{i\pi/2}\left( x-{i\over 2}\pi \right) \prod_{n>0}\left(1+\left(
    {\left(x-{i\over 2}\pi  \right)\over n\pi} \right)^2 \right)
\end{eqnarray}
we get
\begin{eqnarray}
    \prod\limits_{I}{\chi_I\over \tanh \chi_I}
\end{eqnarray}
Finally we integrate $\psi^I_0$ and $x_0^I$  and get the result for
path integral calculation
\begin{eqnarray}
    \tau(M)=Tr\Gamma (-1)^F e^{-\beta H} =\int_M d(Vol) \int
    (d\psi^I_0) \prod\limits_{I} {\chi_I\over \tanh \chi_I}
\end{eqnarray}
which is just Hirzebruch's formula for the signature of the
manifold.

\subsection{Lefschetz fixed-point theorem}
In this section , we use the path integral method to prove the
Lefschetz fixed-point formula. Let $f: M\rightarrow M$ be a smooth
map of a compact oriented manifold into itself. Denote by $H^q(f)$
the induced map on the cohomology $H^q(M)$ . The {\it Lefschetz
number} of f is defined to be
\begin{eqnarray}
    L(f) =\sum\limits_{q}(-1)^q Tr H^q(f)
\end{eqnarray}
At a fixed point P of f the derivative $(Df)_P$ is an endomorphism
of the tangent space $T_PM$ . The {\it multiplicity} of the fixed P
 is defined to be
\begin{eqnarray}
    \sigma_P=\mbox{sgn} \  det ((Df)_P-I)
\end{eqnarray}
Recall that the Lefschetz fixed-point formula states that
\begin{eqnarray}\label{fix formula}
    L(f)=\sum\limits_{P} \sigma_P
\end{eqnarray}

\noindent If f is generated by a vector field on M , (\ref{fix
formula}) can be proved by modifying the action of SQM on M by a
vector field term while still preserving the supersymmtry \cite{AG}.
We show that eqs.(\ref{fix formula}) can also be derived directly
from the standard SQM on Riemannian manifold.

We start again with the lagrangian
\begin{eqnarray}
    L={1 \over 2}g_{IJ}\dot{\phi}^I \dot{\phi}^J +{i\over 2} g_{IJ}(\phi)\bar{\psi}^I D_t \psi^J
    -{i\over 2} g_{IJ}D_t (\phi)\bar{\psi}^I \psi^J
    -{1 \over 4} R_{IJKL}\psi^I \psi^J \bar{\psi}^K \bar{\psi}^L\ \
    \ \ \ \
\end{eqnarray}

We first have a close look at how $H^q(f)$ acts on the state
\begin{eqnarray*}
    H^q(f) \sum\limits_{i_1,\ldots,i_p}
    \varphi_{i_1, \ldots, i_p}(x) dx^{i_1} \wedge \cdots \wedge dx^{i_p} \\
    =\sum\limits_{i_1,\ldots,i_p}
    \varphi_{i_1, \ldots, i_p}(f(x)) df^{i_1}(x) \wedge \cdots \wedge df^{i_1}(x)
\end{eqnarray*}
which just has the effect of sending
\begin{eqnarray}
    &&\phi\longrightarrow f(\phi)\\
    &&\bar{\psi} \longrightarrow f_*(\bar{\psi})
\end{eqnarray}
Recall that $\bar{\psi}$ is regarded as a smooth section of $TM$.\\

So by the standard argument in quantum field theory , we have a path
integral expression for $L(f)$
\begin{eqnarray}
    L(f)&=& Tr H^*(f) (-1)^F e^{-\beta H}\\
        &=& \int_{BC} D\phi D\bar{\psi} D\psi e^{-S_E}
\end{eqnarray}
where the boundary condition is
\begin{eqnarray}\label{bc2}
    &&\phi(\beta)=f(\phi(0))\\
    &&\bar{\psi}(\beta)=f_*(\bar{\psi}(0))\\
    &&\psi(\beta)=\psi(0)
\end{eqnarray}
the Euclidean action is the same as before
\begin{eqnarray*}
    S_E&=&\int_0^{\beta} dt \ \left\{{1 \over 2} g_{IJ}\dot{\phi}^I
    \dot{\phi}^J+{1\over 2}g_{IJ}\bar{\psi}^I D_t \psi^J
    -{1\over 2}g_{IJ}D_t \bar{\psi}^I \psi^J
    +{1 \over 4} R_{IJKL}\psi^I \psi^J \bar{\psi}^K
    \bar{\psi}^L\right\}\label{p1}
\end{eqnarray*}

The path integral still localizes to the constant maps which is just
the constant map to fixed point of f due to the boundary conditions
(\ref{bc2}) . We do the mode expansion respecting the boundary
condition
\begin{eqnarray}
    &&\phi^I=\sqrt{\beta}(tf^I(x_0)+(1-t)x^I_0)+\sqrt{\beta}
    \sum\limits_ {n\neq 0}a_n^I
    e^{2 \pi n i t/\beta}\\
    &&\bar{\psi}^I=\sqrt{2}\left(e^{tA}\right)^I_J \bar{\psi}_0^J +\sum\limits_{n\neq 0}^{\infty} \bar{\psi}_n^I
    e^{2 \pi n i t/\beta}\\
    &&\psi^I=\sqrt{2}\psi_0^I+\sum\limits_{n\neq 0}^{\infty} \psi_n^I
    e^{2 \pi n i t/\beta}
\end{eqnarray}
where formally the matrix A is related to f by
\begin{eqnarray}
    e^{A}=Df(x_0)
\end{eqnarray}
Choose the normal coordinate and take the expansion back to the
action, keeping only the first order, we get
\begin{eqnarray*}
    S_E={1\over 2}x_0^I(\pa_I f(x_0)^K-\delta_I^K)(\pa_K f(x_0)^J-\delta_K^J)
    x_0^J-\bar{\psi}_0^J\left( \left(e^{A}\right)^I_J-\delta^I_J
    \right)\psi_0^I \\+\mbox{excited modes +higer terms}
\end{eqnarray*}
The excited modes of boson and fermion will cancel each other and we
left with the zero modes . The integration of $x_0$ near the fixed
point of f gives
\begin{eqnarray}
    \sqrt{det^{-1}\left( (Df-I)^t (Df-I) \right)}
\end{eqnarray}
the integration of fermion zero modes gives
\begin{eqnarray}
    det\left( Df-I \right)
\end{eqnarray}
All together we have
\begin{eqnarray*}
    L(f)&=&\sum\limits_{q}(-1)^q Tr H^q(f)\\
        &=&Tr H^*(f) (-1)^F e^{-\beta H}\\
        &=& \int_{BC} D\phi D\bar{\psi} D\psi e^{-S_E}\\
        &=&\sum\limits_{P} {det\left( Df|_P-I \right)
        \over \sqrt{det\left( (Df|_P-I)^t (Df|_P-I) \right)}}\\
        &=&\sum\limits_{P} \sigma_P
\end{eqnarray*}

  % 附录
\appendix

\section{Supersymmetric Quantum Mechanics}
\subsection{Witten's definition}
Consider a quantum mechanical system  consisting of a Hilbert (Fock)
space $\it F$ and Hamiltonian H. The system is said to be
supersymmetric quantum mechanical (SQM) if \\

1.{\it F} has a decomposition ${\it F}={\it F}^B \oplus {\it F}^F$
and states in ${\it F}^B$ and ${\it F}^F$ are called bosonic and
fermionic states respectively. There is an operator $(-1)^F$ such
that
\begin{eqnarray}
    &&(-1)^F \Psi =\Psi \ \ if \ \Psi \in {\it F}^B \\
    &&(-1)^F \Psi =-\Psi \ \  if \ \Psi \in {\it F}^F
\end{eqnarray}
F and $(-1)^F$ are called fermion number operator and chirality
operator.\\

2.There are N operators $Q^I$, I=1,$\cdots$,N, such that
\begin{eqnarray}
    Q^I,{Q^I}^\dag &:&{\it F}^B \rightarrow {\it F}^F ,\\
    Q^I,{Q^I}^\dag &:&{\it F}^F \rightarrow {\it F}^B ,\\
    \left\{ (-1^)F,Q^I\right\}&=&\left\{ (-1^)F,{Q^I}^\dag\right\}=0
\end{eqnarray}
$Q^I$ are called supersymmetry(SUSY) charges or generators.\\

3.The SUSY generators satisfy the general superalgebra condition:
\begin{eqnarray}
       \left\{ Q^I,{Q^J}^\dag \right\}&=&2 \delta^{IJ} H\\
       \left\{ Q^I,{Q^J}\right\}&=&\left\{ Q^I,{Q^J}\right\}=0
\end{eqnarray}
where I,J=1,$\cdots$,N.\\
 A quantum system satisfying the above
conditions is said to have a type N supersymmetry.

\subsection{The general structure of Supersymmetric Quantum Mechanics}
To be simple, we assume that the SQM we consider preserve one
supercharge Q. The result for the general case of arbitrary number
of
supercharges is similar.\\
\begin{prpt} The Hamiltonian is a non-negative operator
\end{prpt}
\pf:
\begin{eqnarray}
      H={1 \over 2}\left\{ Q,Q^\dag \right\}\geq0\label{H}
\end{eqnarray}
\ed
\begin{prpt}A state has zero energy if and only if it is annihilated
by Q and $Q^\dag$:
\begin{eqnarray}
    H \Psi=0 \Longleftrightarrow Q \Psi=Q^\dag \Psi=0\ \mbox{for} \
    \Psi\in {\it F}
\end{eqnarray}
\end{prpt}
\pf: \\
if we have
\begin{eqnarray*}
   H \Psi=0 \ \mbox{for} \ \psi\in {\it F}\\
   \Longrightarrow \left( \Psi,H \Psi\right)=0
\end{eqnarray*}
by eqs.(\ref{H}) we have
\begin{eqnarray*}
   &&\left( \Psi,H \Psi\right)\\
   &=&\left( \Psi, \left\{ Q,Q^\dag \right\} \Psi\right)\\
   &=&\left( Q \Psi,Q \Psi\right)+\left( Q^\dag \Psi,Q^\dag \Psi
   \right)=0\\
   \Longrightarrow&&\\
   &&Q\Psi=Q^\dag\Psi=0
\end{eqnarray*}
by the positivity of the inner product on the Hilbert space {\it
F}. The converse case is trivial due to eqs.(\ref{H}).\ed\\ \\
In the physical language, this property is restated as: \emph{The
zero energy ground state is a supersymmetric state and vice versa}.
Thus we also call such a state a \emph{supersymmetric ground state}.

The Hilbert space can be decomposed in terms of eigenspaces of the
Hamiltonian
\begin{eqnarray}
    {\it F}=\bigoplus\limits_{n=0,1,...}{\it F}_{(n)},\ \ H\mid_{{\it
    F}_{(n)}}=E_n.
\end{eqnarray}
We accept the convention that $E_0=0<E_1<E_2<\cdots$. Since Q,
$Q^\dag$ and $(-1)^F$ commute with the Hamiltonian, these operators
preserve the energy levels:
\begin{eqnarray}
    Q,Q^\dag,(-1)^F:{\it F}_{(n)}\longrightarrow {\it F}_{(n)}
\end{eqnarray}
In particular, each energy level ${\it F}_{(n)}$ is decomposed into
even and odd (or bosonic and fermionic) subspaces
\begin{eqnarray}
    {\it F}_{(n)}={\it F}_{(n)}^B\oplus {\it F}_{(n)}^F,
\end{eqnarray}
and the supercharges map one subspace to the other:
\begin{eqnarray}
    Q,Q^\dag:{\it F}_{(n)}^B\longrightarrow {\it F}_{(n)}^F,{\it
    F}_{(n)}^F\longrightarrow {\it F}_{(n)}^B
\end{eqnarray}
Consider the combination $Q_1:=Q+Q^\dag$,which obeys
\begin{eqnarray}
    Q_1^2=2H
\end{eqnarray}
This operator preserves each energy level, mapping ${\it F}_{(n)}^B$
to ${\it F}_{(n)}^F$ and vice versa. Since $Q_1^2=2E_n$ at the nth
level, as long as $E_n>0$, $Q_1$ is invertible and we get the
following
\begin{prpt}
\begin{eqnarray}
    {\it F}_{(n)}^B\cong{\it F}_{(n)}^F\ \ \mbox{for n}
    >0\label{energy}
\end{eqnarray}
\end{prpt}
Note that the bosonic and fermionic supersymmetric ground states do
not have to be paired since at the zero energy level ${\it
F}_{(0)}$, the operator $Q_1$ squares to zero. \emph{Witten index}
is defined to be
\begin{eqnarray}
    Tr(-1)^F e^{-\beta H}=dim {\it F}_{(0)}^B -dim{\it F}_{(0)}^F
\end{eqnarray}
which is independent of $\beta$ by eqs.(\ref{energy}). Physically,
the importance of witten index is that it is invariant under
continuous deformation of the theory(because the states move in
pairs due to the isomorphism eqs.(\ref{energy})) and give a
condition for the broken of supersymmetry.
\\
Now we come to the mathematical structure. Since $Q^2=0 $, we have a
$Z_2$-graded complex of vector spaces
\begin{eqnarray}\label{complex}
  {\it F}^F\xrightarrow{Q}{\it F}^B \xrightarrow{Q}{\it
  F}^F\xrightarrow{Q} {\it F}^B
\end{eqnarray}
and thus we can consider the cohomology of this complex,
\begin{eqnarray}
  H^B(Q):={KerQ:{\it F}^B \rightarrow {\it F}^F \over ImQ:{\it F}^F\rightarrow {\it
  F}^B}\\
  H^F(Q):={KerQ:{\it F}^F \rightarrow {\it F}^B\over ImQ:{\it F}^B\rightarrow {\it
  F}^F}
\end{eqnarray}
The complex in eqs.(\ref{complex}) also decomposes into energy
levels. At the excited level ${\it F}_{(n)}$, we have
\begin{eqnarray}
    \left( Q Q^\dag+ Q^\dag Q\right)/(2E_n)=1
\end{eqnarray}
which implies that the Q-cohomology at the excited level is trivial.
However,at the zero energy level ${\it F}_{(0)}$, the coboundary
operator is trivial, $Q=0$, and we get
\begin{prpt}
\begin{eqnarray}
    H^B(Q)={\it F}^B_{(0)},H^F(Q)={\it F}^F_{(0)}
\end{eqnarray}
\end{prpt}

Finally, we provide the path-integral expression for the Witten
index. By a general physical argument, we have
\begin{prpt}
\begin{eqnarray}\label{path integral}
    Tr(-1)^F=Tr(-1)^F e^{-\beta H}=\int_P DXD\psi D\psi e^{-S_E(X,\psi,\bar{\psi})}
\end{eqnarray}
\end{prpt}

where X denotes the bosonic field and $\psi,\bar{\psi}$ denote the
fermionic field. P on the measure means that we impose the periodic
boundary conditions:
\begin{eqnarray}
    X(\beta)=X(0),\psi(\beta)=\psi(0),\bar{\psi}(\beta)=\bar{\psi}(\beta)
\end{eqnarray}

%\section{a}

%\subsection{section a.1}

%\section{b}

%\subsection{section b.1}

%%%%%%%%%%%%%%%%%%%%%%%%%%%%%%
%% 附件部分
%%%%%%%%%%%%%%%%%%%%%%%%%%%%%%
  % 参考文献
  % 使用 BibTeX
  \bibliography{bib/tex}

\begin{thebibliography}{99}
\bibitem{Atiyah} M.F.Atiyah and I.M.Singer,
 Bull.Am.Math.Soc.69 (1963), 422;\\
M.F.Atiyah and I.M.Singer, Ann, Math.87(1968), 484; ibid., 546.
\bibitem{Morse} E.Witten:{\it Supersymmetry and morse
theory}. J.Diff.Geom.17,661(1982).
\bibitem{AG} L.Alvarez-Gaume,{\it Supersymmetry and the Atiyah-Singer index
theorem},Comm.Math.Phys.90(1983),161.
\bibitem{FW} D.Friedan and P.Windey,{\it Supersymmetric derivation of the Atiyah-Singer index theorem and the
chiral anomaly},Nucl.Phys.B235(1984),395.
\bibitem{Meng} M.Guowu ,{\it A path integral derivation of
$\chi_y$-genus}, J. Phys. A: Math. Gen. 36(2003)
1083-1086,math-ph/0306041.
\bibitem{mirror symmetry} K.Hori, S.Hatz, A.Klemm, R.Pandharipande,
R.Thomas, C.Vafa, R.Vakil and E.Zaslow, "Mirrow Symmetry". Clay
Mathematics Monographs 1, American Mathematical Societry,
Providence, Clay Mathematics Institute, Cambridge, 2003, ISBN
0-8218-2955-6.
\bibitem{chen} S.S.Chern , On the curvature integrals in a
Riemannian manifold. Ann. Math. 46 , 674(1942)
\end{thebibliography}

\end{document}